\documentclass{raa}
\usepackage{graphicx,times}
\usepackage{amssymb,amsmath}

\usepackage[a4paper=true,dvipdfm=true,pagebackref=true]{hyperref}
\hypersetup{pdftitle = The title of my PDF, pdfauthor = My name, pdfsubject= The subject, pdfkeywords = keyword1 keyword2 keyword3}
\hypersetup{colorlinks = true, linkcolor = green, anchorcolor = red, citecolor = blue, filecolor = red, pagecolor = red, urlcolor = red}

\newcommand{\teffa} {\mbox{\rm $Teff_{APOGEE}$}}
\newcommand{\logga} {\mbox{\rm $logg_{APOGEE}$}}
\newcommand{\feha} {\mbox{\rm $[Fe/H]_{APOGEE}$}}
\newcommand{\teffl} {\mbox{\rm $Teff_{LAMOST}$}}
\newcommand{\loggl} {\mbox{\rm $logg_{LAMOST}$}}
\newcommand{\loggs} {\mbox{\rm $logg_{SVM}$}}
\newcommand{\fehl} {\mbox{\rm $[Fe/H]_{LAMOST}$}}
\newcommand{\feh} {\mbox{\rm [Fe/H]}}
\newcommand{\teff} {\mbox{\rm $T_{eff}$}}
\newcommand{\logg} {\mbox{{\rm logg}}}

\begin{document}


\title{
The Comparison of Stellar Atomspheric Parameters between LAMOST and APOGEE
databases} 
\volnopage{ {\bf 2015} Vol.\ {\bf X} No. {\bf XX}, 000--000}
   \setcounter{page}{1}

\author{Y.Q. Chen\inst{1}, G. Zhao\inst{1}, C. Liu\inst{1}, J. Ren\inst{1}, Y.P. Jia\inst{1}, J.K. Zhao\inst{1}, A.L. Luo\inst{1}, Y. Zhang\inst{2}, Y.H. Hou\inst{2}, Y.F. Wang\inst{2} \& M. Yang\inst{2}}
\institute{Key Laboratory of Optical Astronomy, National Astronomical Observatories, Chinese Academy of Sciences, Beijing, 100012, China; {\it cyq@bao.ac.cn} \\ \and
Nanjing Institute of Astronomical Optics \& Technology, National Astronomical Observatories, Chinese Academy of Sciences, Nanjing 210042, China\\}


\date{Received~~2015 month day; accepted~~2015~~month day}

\abstract{We have compared the stellar parameters, temperature, gravity and metallicity,
between the LAMOST-DR2 and SDSS-DR12/APOGEE database for stars in common.
It is found that the LAMOST database provides a better red-clump feature
than the APOGEE database in the $\teff$ versus $\logg$ diagram.
With this advantage, we have separated red clump stars from
red giant stars, and attempt to establish the calibrations between
the two datasets for the two groups of stars respectively.\\
It shows that there is a good consistency in temperature with a calibration 
close to the one-to-one line, and we can establish a satisfied
metallicity calibration of $\feha=1.18\fehl+0.11$ with
a scatter of $\sim0.08$ dex for both red clump and red giant branch samples.
For gravity, there is no any correlation for red clump stars between the two databases, and
scatters around the calibrations of red giant stars are substantial.
We found two main sources of the scatters of  $\logg$ for red giant stars.
One is a group of stars with $0.00253*\teff-8.67<\logg<2.6$
locating at the forbidden region, and the other is the contaminated red clump stars, which
could be picked out from the unmatched region where stellar metallicity is
not consistent with the position
in the $\teff$ versus $\logg$ diagram. 
After excluding stars at the two regions,
we have established two calibrations for red giant stars,  $\logga=0.000615*\teffl+0.697*\loggl-2.208$ ($\sigma=0.150$)
for $\feh>-1$ and $\logga=0.000874*\teffl+0.588*\loggl-3.117$ ($\sigma=0.167$) for $\feh<-1$.
The calibrations are valid for stars with $\teff=3800-5400\,K$
and $\logg=0-3.8$ dex, and are useful in a work aiming to combine
the LAMOST and APOGEE databases in the future study.
In addition, we find that an SVM method based on seismic $\logg$ is a good way
to greatly improve the accuracy
of gravity for these two regions at least in the LAMOST database.
}

\authorrunning{Y.Q. Chen et al. }            
\titlerunning{Stellar Parameters between LAMOST and APOGEE databases}                    
\maketitle

\section{Introduction}
Stellar spectroscopic survey provides an important source
of knowledge in astrophysics. The information from spectroscopy
include physical parameters of stars (temperatures, gravities
and detailed chemical composition) and their kinematics
(radial velocities), which are crucial to our understanding of
stars and stellar populations in the Milky Way and other galaxies.
The advent of large stellar spectroscopic surveys like SEGUE/SDSS
(Yanny et al. 2009), RAVE (Kordopatis et al. 2013),
APOGEE (Majewski et al. 2015), and the LAMOST (Zhao et al. 2012)
is leading Galactic astronomy into a precision science, where we can
identify different sub-populations by combining the chemical
composition with kinematics and
trace Galactic evolution and stellar structure at various Galactic
locations in details.

Recently, the LAMOST telescope,  also known as the Wang-Su Reflecting
Schmidt Telescope or the Guoshoujing Telescope (Cui et al. 2012), has finished
a two year regular survey after one year's pilot survey
in 2011. The combination of a large
aperture (4 m) and high multiplex factor (4000 fibers)
with a 5 degree field of view makes it a unique facility.
With the low-resolution (R = 1800), the LAMOST project currently
presents spectra of $\sim 4,136,000$ objects and stellar parameters
for  $\sim 2,200,000$ stars in its second data release (Luo et al. 2015).
This database includes many previously observed Kepler targets 
provided by the LAMOST-Kepler project (Cat et al. 2014).
More detailed information on the data release can be found in
the website (http://www.lamost.org/public/).
Stellar parameters in the LAMOST DR2 database are derived by
the package $ULySS$ (Wu et al. 2011), where an observed spectrum is fitted
against a model expressed as a linear combination of non-linear components,
optionally convolved with a line-of-sight velocity distribution (LOSVD)
and multiplied with a polynomial function.

Coincidently, 
APOGEE (Majewski et al. 2015) has released its
three-year, near-infrared survey for 100,000 red giant stars
included as part of the SDSS-III (Eisenstein et al. 2011;Ahn et al. 2012).
With a resolving power of 22,500, APOGEE is able to derive detailed
abundances for 15 elements as well as the three basic
stellar parameters, temperatures, gravities and metallicities.
Based on a synthetic grid of Kurucz models and an efficient search method,
a best match within the synthetic grid is found for each APOGEE spectrum
to provide the initial set of parameters: Teff, log g,
[M/H], [C/M], [N/M], and [$\alpha$/M]. Then the stellar parameters
are calibrated by giants in the Kepler field and stars in the
clusters.

The main population of the observed targets in both
the LAMOST and APOGEE surveys is the Galactic disk. In principle, they
can be merged together to probe the property of the Galactic disk, and
the results obtained from either one can be checked in an independent way.
Meanwhile, the two surveys are quite different in may ways. They
have different observed bands, resolving powers of the spectra,
and thus they can provide abundances for different elements and kinematics
with different precisions.
The two surveys have their own advantages in the sky coverage, selection function
and stellar spectral type. Thus, it is important to mix these information
together in order to understand the history of the Galactic disk
from different views as well as checking if the results from either survey
are reliable or not. In view of this, we aim to make a systematical
study on the chemical and kinematic properties of the Galactic disk
by combining the data from the two surveys in the near future. This combination is
particularly important for the study of the local effect of the chemical
evolution, stellar migration of the Galactic disk and 
the understanding of the origins of many kinematical structures in the Galactic disk.

It is interesting to know how consistent stellar parameters between 
the LAMOST and APOGEE databases are, and
if it is possible to establish some kinds of transformation relations for stellar
parameters between the two databases so that they can be combined in the
future study on the Galaxy. Since the APOGEE database is based on high
resolution spectra with high signal-to-noise ratios, it may provide
better parameters, at least for stellar metallicity, than the LAMOST
database estimated from low resolution and low signal-to-noise spectra.
Meanwhile, we hope to check the stellar parameters provided by the LAMOST DR2 database
before they can be really used to probe the evolution of the Galaxy.
Specifically, we want to know what kinds of stars in the LAMOST DR2 database
have reliable parameters, and what kinds of stars show unreasonable
values which should be excluded in the future Galactic study.
Finally, this check and comparison might provide
some clues too improve the stellar parameters in both LAMOST DR2 and
APOGEE databases.
In this work, we aim to select a sample of
common stars with high quality spectra in both databases,
compare stellar parameters and establish calibrations for individual
stellar parameters if possible.

\section{Star Sample and its Division into RC and RGB Subsamples}
The selection procedure of the sample stars is proceeded as follows.
First, we select common stars with the same Galactic locations,
(RA,DEC) within 3 arcsecs, in the LAMOST DR2 and SDSS DR12/APOGEE
databases. Then we limit stars with stellar parameters in the
reasonable regions of $3000<\teff<9000\,K$, $-1.0<\logg<6.0$ and
$-5.0<\feh<1.0$. Third, we select stars with high signal-to-noise
spectra in both surveys, the signal-to-noise in $g$ band
of LAMOST spectra $sng > 30$ and the signal-to-noise of APOGEE spectra
$sn>100$. With these criteria, we have 5626 stars in common, for which 
the $\teff$ versus $\logg$ diagrams are shown in Fig.~1.
We notice that there are some turn-off and subgiant stars.
Since the APOGEE database only provides stellar
parameters for giants, we thus limit our sample of stars with $\logg<3.8$ in both
database. In total, we have 5352 stars for comparison.

There are some differences in the $\teff$ versus $\logg$ diagrams
between the two datasets. The most prominent difference is that 
the LAMOST database shows a strong clump feature in this diagram which corresponds
to the well-known red clump population, which is not easily seen
by eye check in the APOGEE database. The appearance of this feature proves that stellar
parameters in the LAMOST database are generally reasonable at least
for stars at this region. With this advantage, we select a sample of RC stars from the
LAMOST database by limiting stars within the two red lines, where stars follow
the relation of $-0.0010*\teffl+7.10 <\loggl<-0.0005*\teffl+5.05$, and
in the temperature range of $4600<\teffl<5000\,K$.
The right temperature limit of $\teffl<5000\,K$ is applied since the star number
significantly decreases in the lower panel of Fig.~1, and this criterion
may exclude the secondary red clump sequence, which is predicted to locate
on the blue and faint-magnitude side of the main red clump sequence in
the color magnitude diagram (Girardi et al. 1999).
The left temperature limit of $\teffl>4600\,K$ is chosen in order to avoid the
contamination of the possible bump of red giant branch at the red
side of red clump at the solar metallicity.
For comparison, two theoretical isochrones of 1 Gyr and 10 Gyr with Z=0.030
from the Padova website (http://stev.oapd.inaf.it/cgi-bin/cmd,
Bressan et al. 2012) are overploted in Fig.~1.

The selection procedure of RC sample is mainly based on eye check
on the $\teff$ versus $\logg$ diagram of the LAMOST database.
According to \cite{Bovy14}, RC stars in the
APOGEE database can be selected by their position in color-
metallicity-gravity-temperature space 
using a new method calibrated using stellar evolution
models and high-quality asteroseismology data. In their
Figure 1, a linear line 
of $\logga=0.0018*(\teffa-4607)+2.5$ at solar metallicity
clearly separates RC stars from RGB stars,
which is shown in red line in the upper panel of Fig.~1.
When we overplot our selected RC sample of stars with green dots 
in the $\teff$ versus $\logg$ diagram of the APOGEE database,
they locate exactly on the left edge of the red line.
Thus, our RC sample generally follows the selection criteria of \cite{Bovy14}.
Note that our RC sample stars do not take into account stars on the secondary red clump sequence
for two reasons. First, they can be identified by seimic analysis (Stello et al. 2013)
but it is difficult to pick them out from the $\teff$ versus $\logg$ diagram. Secondly,
they might have different properties as compared with stars in the main RC sequence; they are
massive, young and have different dependences of $\logg$ with $\teff$.
Thus, this work mainly involve with the main sequence RC stars.
Finally, we divide our selected sample of 5352 stars
into two samples, the RC sample
with 1544 stars and the RGB sample from the remaining 3808 stars.
In the following sections, the two samples will be investigated {separately}
due to their different properties.

\begin{figure}[ht]
\includegraphics[width=10.5cm, angle=0]{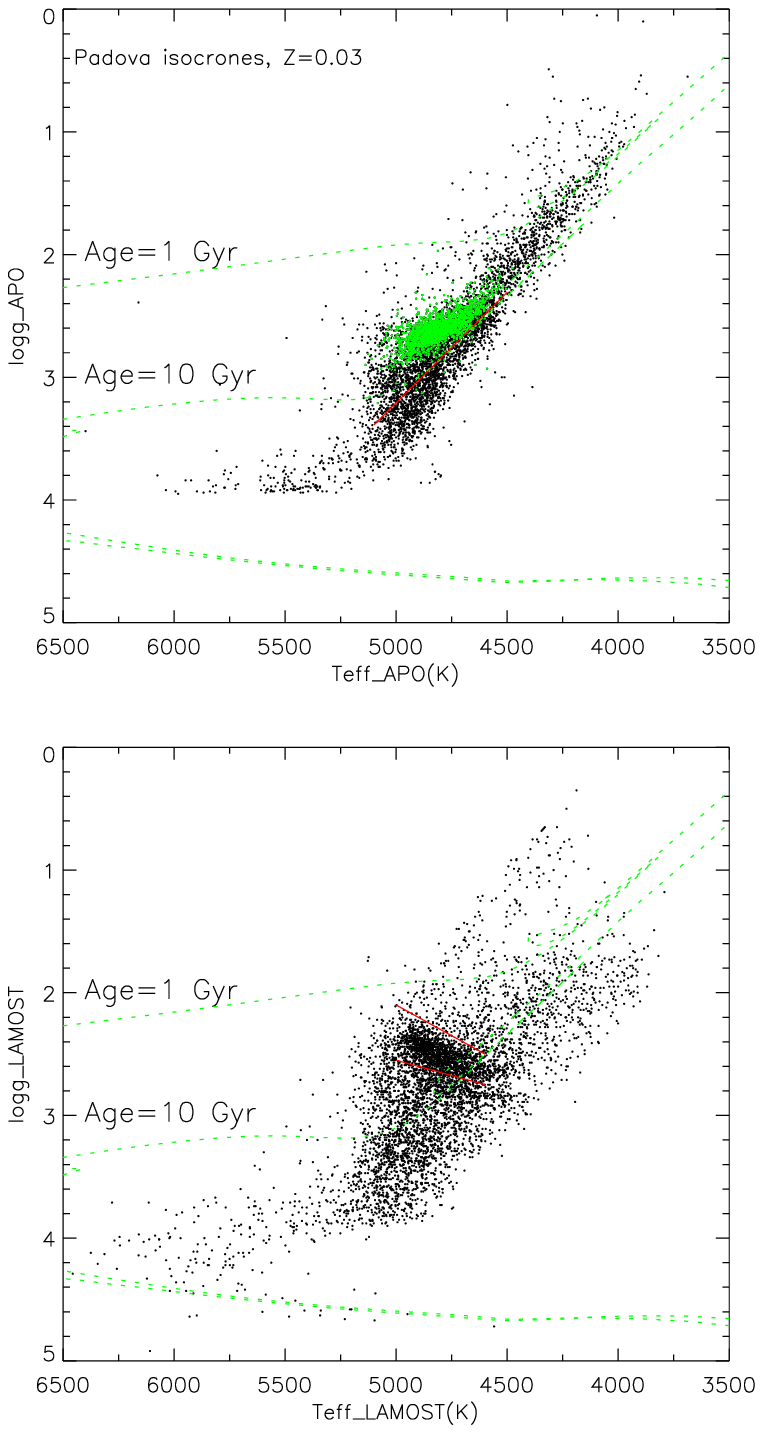}
\caption{The $\teff$ versus $\logg$ diagrams for the LAMOST (lower panel)
and APOGEE (upper panel) databases based on 5626 common stars with
high quality spectra (LAMOST:$sng>30$, APOGEE:$sn>100$). Dashed lines show
theoretical isochrones of 1 and 10 Gyr at Z=0.030
from the Padova website (Bressan et al. 2012). The selection
criteria of RC stars are marked in red lines, and green dots
in the upper panel are our sample of RC stars selected from the LAMOST database.}
\label{fig1}
\end{figure}

\section{The Comparsion of Stellar Parameters between the LAMOST and APOGEE databases}

\subsection{The $\teff$, $\logg$ and $\feh$ Distrbutions}
In order to investigate if there is any systematic shifts in stellar parameters
between the LAMOST and APOGEE databases, we show the distributions
of $\teff$, $\logg$ and $\feh$ in Fig.~2 for the whole sample.
It shows that 
there are systematic shifts in the $\logg$ and $\feh$ distributions
but not clear difference for the $\teff$ distribution between the two datasets.
For gravity, the LAMOST dataset shows one peak at $\logg\sim2.3-2.4$, while
the APOGEE dataset has the main peak at $\logg\sim2.5-2.6$ and possible
second peak at $\logg\sim2.85$, which may correspond to the secondary
red clump sequence according to \cite{Bovy14}. For metallicity, there
is a prominent shift in the metallicity peak from $\feh\sim-0.1$ in the APOGEE dataset 
to $\feh\sim-0.3$ in the LAMOST dataset as well as the whole shift of the distributions
at the metal rich side. In addition, most stars
have metallicity in the range of $-1.0<\feh<0.5$ indicating the
dominate population of the Galactic disk in our sample.
Note that the adopted $\logg$ and $\feh$ values in the APOGEE database
have been corrected by equations (3) and (6) of \cite{Holtzman15}, without which
the differences between the two datasets will be even larger.

\begin{figure}[ht]
\includegraphics[width=10.5cm, angle=0]{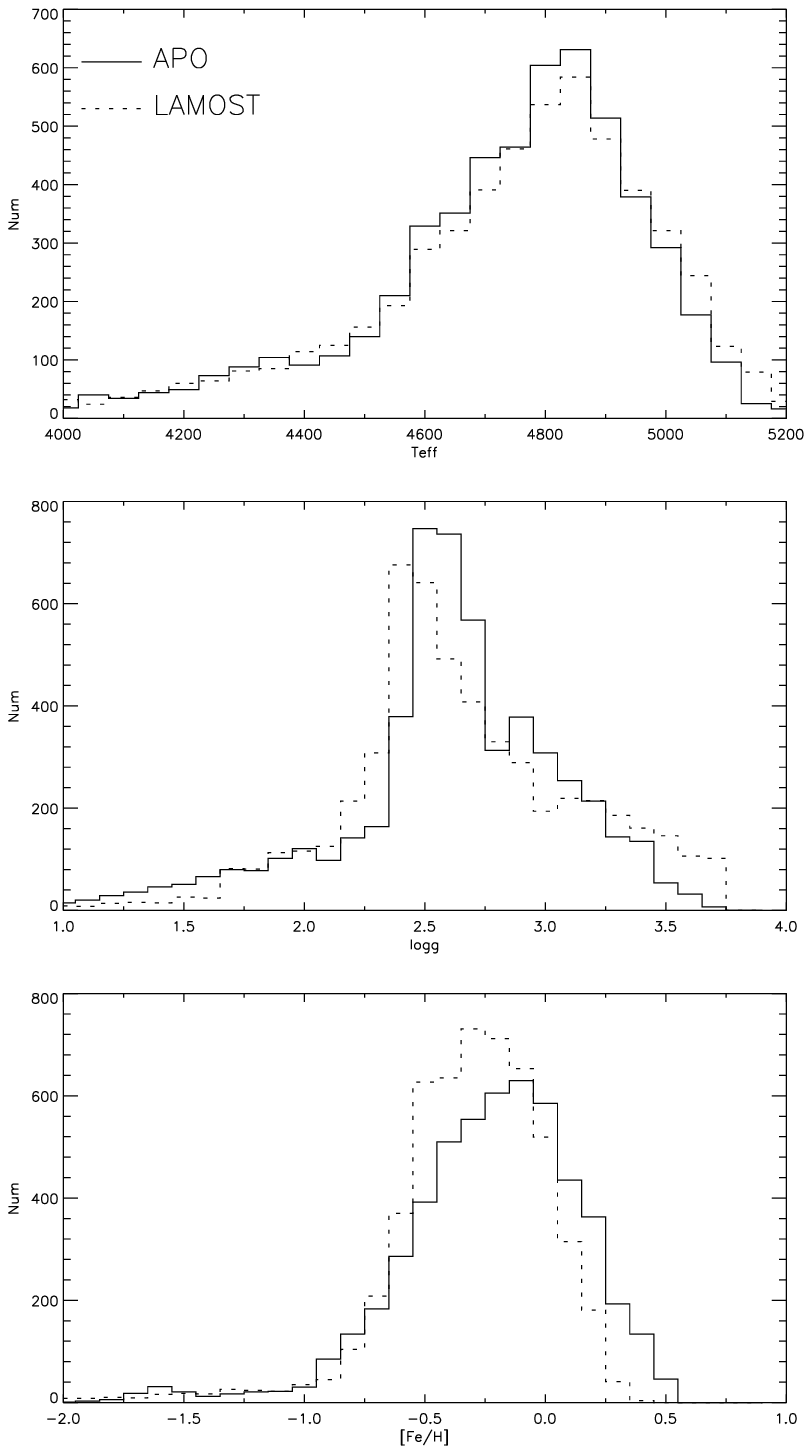}
\caption{The distributions of $\teff$, $\logg$ and $\feh$ for
the whole sample based on the LAMOST (dash lines)
and APOGEE (solid lines) databases.}
\end{figure}

\begin{figure*}[ht]
\includegraphics[width=14.5cm, angle=0]{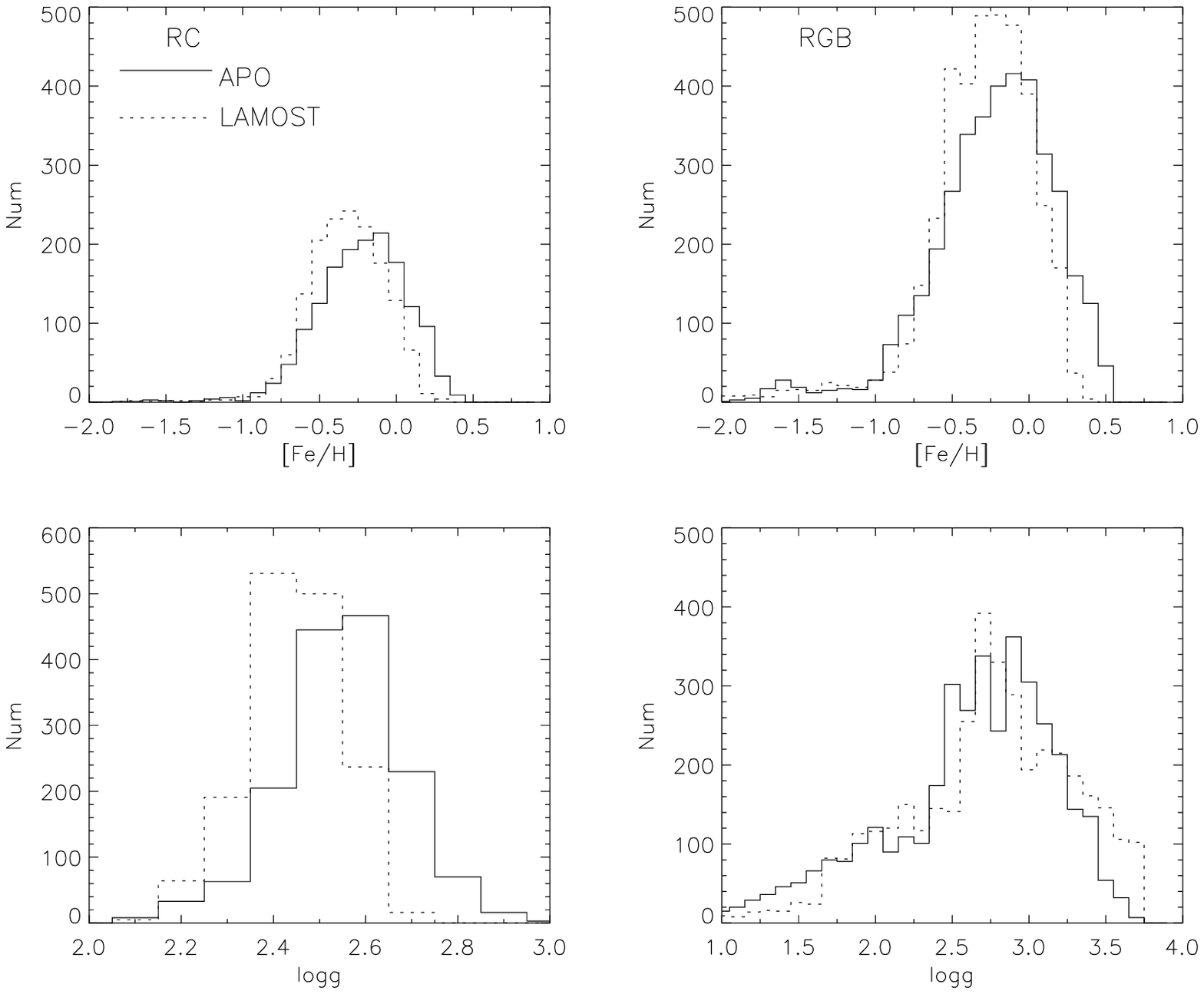}
\caption{The distributions of $\logg$ and $\feh$ for
RC (left)  and RGB (right) stars based on the LAMOST (dash lines)
and APOGEE (solid lines) databases.}
\end{figure*}

Fig.~3 shows the $\feh$ and $\logg$ distributions for RC and RGB samples,respectively.
Clearly, the metallicity shift is systematic, and RC and RGB samples
behave in a similar way. The systematic shift of gravity in Fig.~2 mainly comes
from the RC sample, and it is not so prominent for RGB sample. Instead, the APOGEE
dataset shows a slightly broader $\logg$ distribution than the LAMOST dataset.
Moreover, as shown in Fig.~1, the dependence of $\logg$
with $\teff$ in the RC sample is opposite between the LAMOST and
APOGEE dataset. In view of these different properties for RC and RGB samples,
it is reasonable to establish the calibrations for RC and RGB stars separatedly.

\subsection{The Comparsion and Calibrations of Stellar Parameters}
If possible, we aim to establish the calibrations of stellar parameters
between the LAMOST and APOGEE datasets in order to combine the two
surveys in the future study. For this purpose,
the one-to-one comparisons of $\teff$, $\logg$ and $\feh$ 
for RC (left panels) and RGB (right panels) samples
between the LAMOST and APOGEE databases are shown in Fig.~4.
Stars with  $\feh<-1$ are marked by red crosses since they only
contribute small parts of our main population of the Galactic 
disk with $\feh>-1$ (see Fig.~2).

Note that most stars in our RC sample have $\feh>-0.8$ which is
consistent with the metallicity distribution of local RC stars as
shown in \cite{Puzeras10}. However, 22 stars in our RC sample
have $\feh<-1$, which is outside the metallicity range
of local RC sample of $-0.8<\feh<0.3$ (their Figure 5).
We marked these RC stars with  $\feh<-1$ by red crosses
in the left panels of Fig.~4. We find that they do not follow
the general trends of most RC stars in both $\teff$ and $\logg$
panels of Fig.~4. Thus, they might not be real RC stars. Instead,
they could be red horizontal branch or metal poor red giant stars, but it is
difficult for us to distinguish them.
Thus, these stars are excluded in the following analyses.
For RC stars, the agreement in $\teff$ between the LAMOST and APOGEE
datasets is good with a std scatter of 53\,K around the
calibration of $\teffa=0.95*\fehl+210$.
Note that significant deviations in the temperature comparison
from the one-to-one line at both ends mainly come from the 
selection criterion of $4600<\teffl<5000$\,K for RC stars.
However, there is no any correlation or any kind
of anti-correlation in the $\logg$ comparison, and
the scatters are too large to obtain
reliable calibrations.
For metallicity, a good calibration of $\feha=1.16*\fehl+0.14$ with a scatter
of 0.065 dex can be established for RC stars with  $\feh>-1$.

For RGB stars, a similar metallicity calibration
of $\feha=1.18*\fehl+0.11$ ($\sigma =0.08$) as RC sample
is found, and we may establish a $\teff$ calibration of $\teffa=0.86*\teffl+665$ with
a scatter of 77\,K. For gravity, the general trend in the comparison
follows the one-to-one line, but there are very large scatters in the
range of $1.5<\loggl<2.6$. Moreover, stars with  $\feh<-1$, marked
by red crosses again, have systematically higher value by $\sim0.2$ dex
for stars with $\loggl>1.0$.
In view of this difference, we establish two gravity calibrations for 
RGB stars with a metallicity division at $\feh=-1$,
$\logga=0.92*\loggl+0.09$ for $\feh>-1$
and $\logga=0.90*\loggl+0.34$ for $\feh<-1$, with a scatter of $0.24$ in
both calibrations. When the temperature term is included, the scatters
are slightly reduced with calibrations of
$\logga=0.00105*\teffl+0.46*\loggl-3.85$ ($\sigma=0.20$)
for  $\feh<-1$  and of $\logga=0.00087*\teffl+0.59*\loggl-3.11$ ($\sigma=0.17$)
for $\feh>-1$. As we further include the metallicity term,
the calibrations are of
$\logga=0.00140*\teffl+0.28*\loggl+0.20*\fehl-4.89$ ($\sigma=0.20$)
for  $\feh<-1$  and of $\logga=0.00107*\teffl+0.46*\loggl+0.34*\fehl-3.64$ ($\sigma=0.15$)
for $\feh>-1$.

\begin{figure*}[ht]
\includegraphics[width=14.5cm, angle=0]{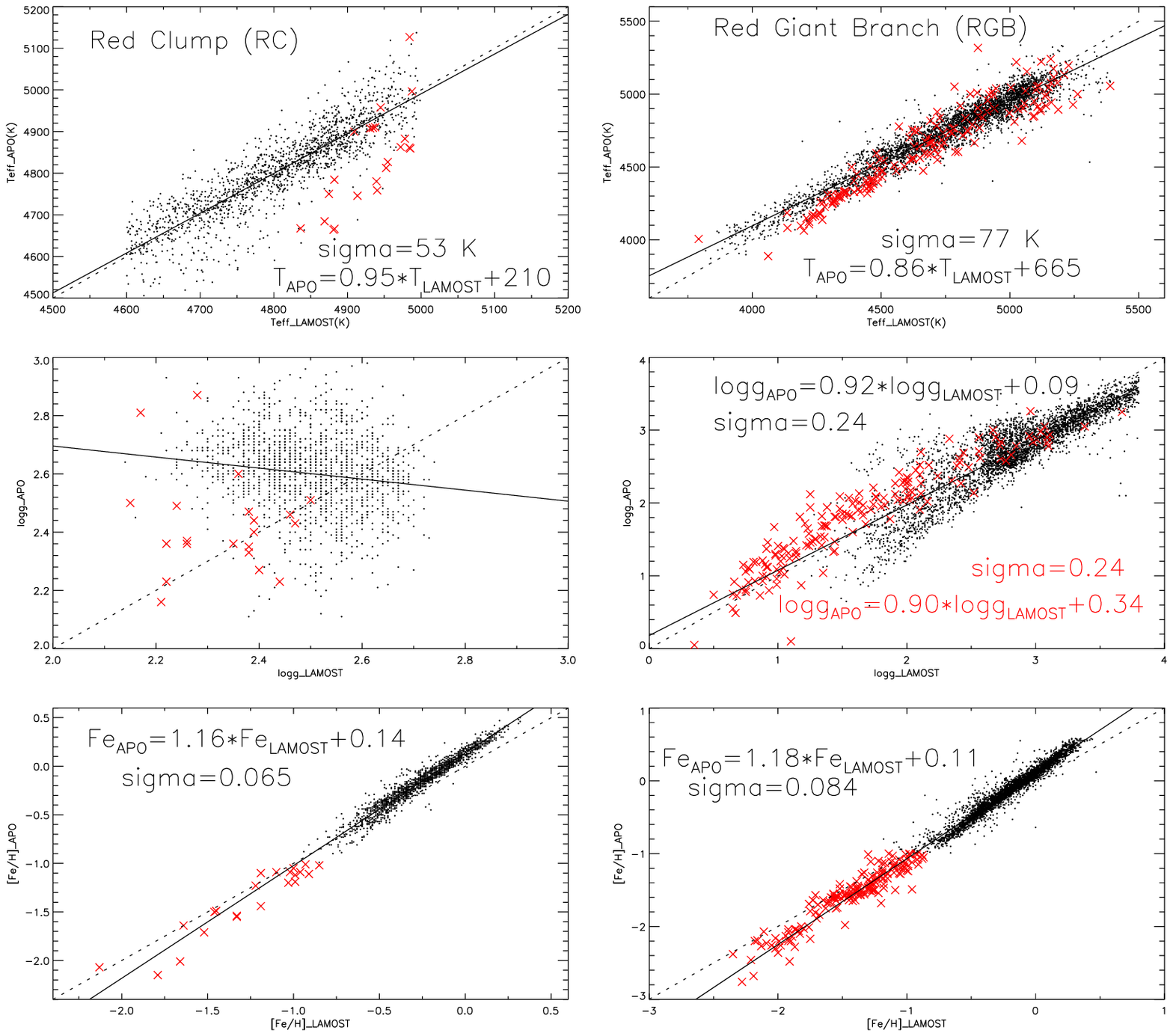}
\caption{The comparisons and calibrations of $\teff$, $\logg$ and $\feh$ between
the LAMOST and APOGEE databases for RC (left panels) and RGB (right panels) samples.
Dashed lines are the one-to-one relations while solid lines are the calibrations.
Stars with $\feh<-1$ are indicated by red crosses.}
\end{figure*}

\subsection{Refine the $\logg$ calibrations for RGB stars with $\feh>-1$}
In order to probe the large scatter in the $\logg$ calibration between the APOGEE and the LAMOST
datasets for RGB stars, we carefully inspect their differences in the $\teff$ versus $\logg$
diagrams in Fig.~5. It shows that the main discrepancy comes from the lack of stars
on the right side of the blue dash line, which can be
expressed by the relation of $\logg=0.00253*\teff-8.67$ by passing the two points
in ($\teff$,$\logg$) of (3500,0.2) and (5000,4.0).
The solid line in the left-upper panel of Fig.~5
shows an extreme case with an isochrone of 16 Gyr at the metallicity of $Z=0.030$ 
from Padova group (Bressan et al. 2012), and we find substantial number of stars
in the LAMOST dataset locate on the right side of this extreme case.
We thus define a forbidden region where no theoretical model could predict 
such kind of stellar parameters.
Taking into account of temperature and gravity uncertainties
as well as the theoretical isochrone of 16 Gyr at $Z=0.030$,
we may assign stars with $\loggl<2.6 $ in the LAMOST dataset and locating on the right side 
of blue line belonging to a forbidden region; 
they are marked by blue dots in all panels of Fig.~5.
Obviously, these blue dots locate below the one-to-one line in the comparison of $\logg$
in the left-lower panel of Fig.~5, indicating the LAMOST values are too high as compared
with the APOGEE values. These stars become one of the main sources
of the  $\logg$ scatter in the comparison, and thus they
should be excluded in the calibrations.

Meanwhile, recall that our selection criteria of RC stars are quite strict
in order to obtain a clear sample, and thus our RGB sample from the remaining 
stars is somewhat contaminated by some RC stars.
In particular, the secondary RC stars if existed, locate at the blue side of the main
RC sequence, are included in our RGB sample. We notice that these stars
can be distinguished by matching their locations in the $\teff$ versus $\logg$ diagram
with their metallicities in the sense that metal poor RGB stars with $\feh<-1$
will locate at the blue side of the green dash line, which arbitrarily shift
the blue line by 400\,K in temperature. We do not adopt
the theoretical isochrone of 16 Gyr at the metallicity of $Z=0.030$
from Padova group (Bressan et al. 2012) because they do not fit the
LAMOST dataset statistically. But we have checked that the shift of 400\,K in temperature
corresponds to the change of RGB ridge lines from solar metallicity to $\feh=-1$.
Specifically, stars with $\feh>-1$
but locating on the blue side of the green dash line could be
RC stars instead of RGB stars or RGB stars with wrong stellar parameters. 
These stars are marked by green dots in Fig.~5,
and they constitute the second main source of scatter in the $\logg$ comparison.
Generally, they have higher $\logg$ values in the APOGEE dataset than those of 
the LAMOST databases. These stars are further excluded in the calibration.

After excluding stars from the above two regions, we repeat the procedures
and obtain the calibration of $\logga=0.000615*\teffl+0.697*\loggl-2.208 (\sigma=0.150)$
for RGB stars with $\feh>-1$.
The same procedures are performed from the APOGEE dataset to the
LAMOST dataset, and we obtain the 
calibration of $\loggl=-0.000941*\teffa+1.344*\logga+3.674 (\sigma=0.167)$
for  $\feh>-1$.

\begin{figure*}[ht]
\includegraphics[width=14.5cm, angle=0]{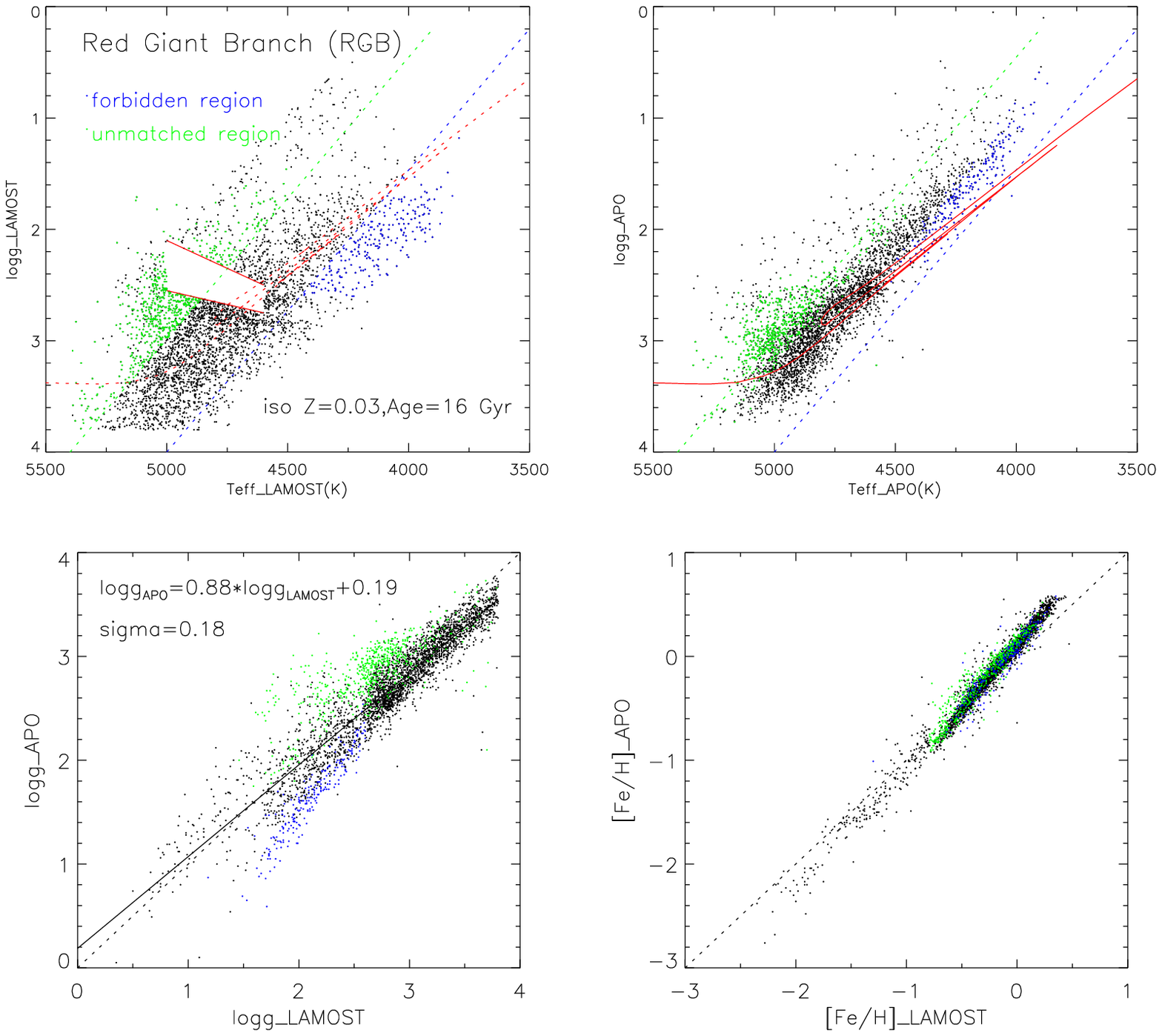}
\caption{Upper: The $\teff$ versus $\logg$ diagrams for RGB stars in
the LAMOST and APOGEE datasets. Blue dash line is passing two points
at ($\teff$,$\logg$) of (3500,0.2) and (5000,4.0) and red dash line is
the sochrone of 16 Gyr at $Z=0.030$ from Padova group (Bressan et al. 2012).
Lower: The comparison of gravity and metallicity for different groups
of RGB stars.
Stars with $2.6>\logg>0.00253*\teff-8.67$ at the forbidden region
are marked by blue dots and stars at the unmatched region are marked
by green dots.}
\end{figure*}

\subsection{On the Gravity Discrepancy of RC stars}
As described in \cite{Holtzman15}, RC stars in the APOGEE database
are calibrated as the same equation (Eq. 3) as RGB stars. This is not
the best solution for RC stars. Instead, the comparsion of $\logg$ between
the raw ASPCAP values (Garcia et al. 2015) and the seismic ones from the APOKASC
catalog (Pinsonneault et al. 2014) in the Kepler field indicates
a difference of 0.15 dex between RC and RGB stars (Holtzman et al. 2015). That means we
need to reduce $\logg$ in the APOGEE database by a further order
of 0.15 dex for RC stars. If this difference is applied, the
$\logg$ distributions between the LAMOST and APOGEE databases
in the left-lower panel of Fig.~3 are consistent. This consistency
shows that the LAMOST $\logg$ for RC stars is on the same scale
as that of the seismic values from the Kepler survey.

The second discrepancy for RC stars between the LAMOST and the APOGEE databases
(see Fig.~1) is the opposite dependence of $\logg$ on $\teff$. In the LAMOST dataset,
$\logg$ increases with decreasing $\teff$ with a slope of $-0.68$ dex per 1000\,K, while the slope
is of $0.89$ dex per 1000\,K in the APOGEE dataset. This discrepancy is the same as we
limit stars with $\feh>-0.5$ in both samples. Since the selection of RC stars
is carried out in the LAMOST dataset and we limit stars in the LAMOST temperature range of
$4600<\teff<5000$\,K, the slope of $-0.68$ dex per 1000\,K just reflects
our selection criteria of $-0.0010*\teffl+7.10 <\loggl<-0.0005*\teffl+5.05$. Moreover,
the slope will be reduced after excluding a few stars at $\teff\sim4980$\,K and $\logg\sim2.1$, 
and the scatter of $0.10$ dex at a given $\teff$ will significantly affect this slope.
However, there is a strong dependence of $\logg$ on $\teff$ in the APOGEE database,
which can not be explained by its scatter. The strong dependence of $\logg$ on $\teff$ persists
even the systematic shift of 0.15 dex is applied 
to RC stars in the APOGEE database.

Since our sample of RC stars fits well the selection criteria of \cite{Bovy14},
we can expect that they are real RC stars in both databases. With this agreement, we
plot an independent sample of RC stars from \cite{Casagrande14} by red open circle
in  Fig.~6
and compare the dependence trends of $\logg$ on $\teff$ among the three datasets.
Note that the RC sample in \cite{Casagrande14} has several advantages.
(i) RC stars are identified by seismic data with period space from \cite{Stello13}
and they have accurate seismic $\logg$.
(ii) They have str\"omgren $(b-y)_0$ colors, which are very sensitive to temperature.
(iii) \cite{Casagrande14} provide mass and age for RC stars, which allow
us to limit stars with $mass<1.8 M_{\odot}$
and $age>2$ Gyr in order to exclude the RC stars on the secondary sequence.
Clearly, there is no slope in the $\teff$ versus  $\logg$ diagram, and most
stars have $\logg=2.4-2.6$. The LAMOST data agree with
\cite{Casagrande14} for RC stars in a better way than the APOGEE database.
 Careful inspection of
the Figure 4 of \cite{Holtzman15}, there is a hint of increasing trend of $\Delta logg (ASPCAP-Kelper)$
with increasing $logg_{ASPCAP}$ for RC stars (blue squares). If this trend is
applied to the APOGEE database, the slope in the $\teff$ versus  $\logg$ diagram
for RC stars in the APOGEE databases will be slightly reduced. But further work on
this correction should be done in the future.

\begin{figure}[ht]
\includegraphics[width=14.5cm, angle=0]{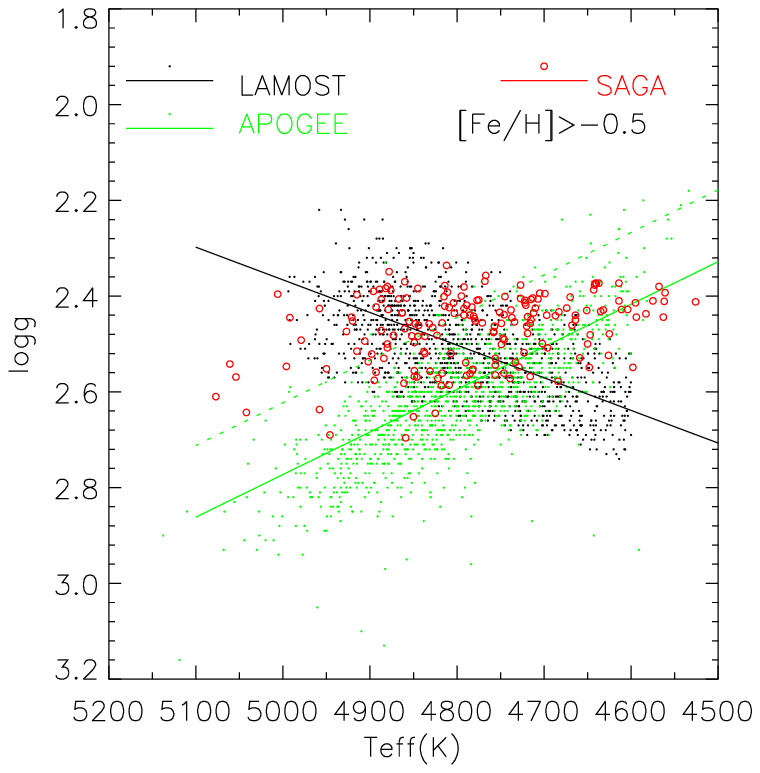}
\caption{The comparison of the $\teff$ versus $\logg$ diagram for RC stars
in the LAMOST and APOGEE datasets with that from the SAGA survey by \cite{Casagrande14}.}
\end{figure}

\subsection{New LAMOST Gravities from SVM with Seismic Basement}
As described above, the LAMOST database do not provide
the best gravities for giant stars in the forbidden region and
the slope in the $\teff$ versus  $\logg$ diagram
for RC stars is not consistent with that from the SAGA survey by \cite{Casagrande14}.
Is there some way to improve these gravities? Recently, \cite{Liu15}
present a support vector machine (hereafter SVM) method to derive gravities for
LAMOST giants based on a sample of stars with seismic $\logg$ in \cite{Huber14}
as a training dataset. In this training dataset,
$\logg$ values in \cite{Huber14} has been {re-calculated} with
the LAMOST  $\teff$, and thus these new gravities
in \cite{Liu15} match $\teff$ values in the LAMOST database used in the present work.

It is interesting to investigate how stars in the forbidden
and unmatched regions in Fig.~5 behave with these new gravities. Fig.~7
shows the  $\teff$ versus  $\loggs$ diagram for 4915 stars.
Interestingly, most stars in the forbidden region (blue dots) locate
around the blue dash line, on the right side of which the APOGEE database 
is lack of stars. Clearly, new gravities from \cite{Liu15} seem to be more
reasonable and they are consistent with those in the APOGEE database 
in this region within the errors. Meanwhile, a substantial number of 
stars in the unmatched region in Fig.~5 locate at the left part of the
main RC stars, indicating that they are probably also main RC stars instead of
secondary ones. In addition, the main feature of RC stars becomes quite prominent
in the $\teff$ versus  $\loggs$ diagram as stars in the unmatched region of
in Fig.~5 are included. However, the selection of stars within the two red lines
may not be suitable based on new gravities. Instead, there is not significant dependence
of $\logg$ with $\teff$ for the main RC feature. In Fig.~8, we plot the $\teff$ 
versus $\loggs$ diagram for our selected RC sample of stars with $\feh>-0.5$. The slope
is of $0.18$ dex per 1000\,K which is consistent with that from the 
SAGA survey by \cite{Casagrande14}.

\begin{figure}[ht]
\includegraphics[width=14.5cm, angle=0]{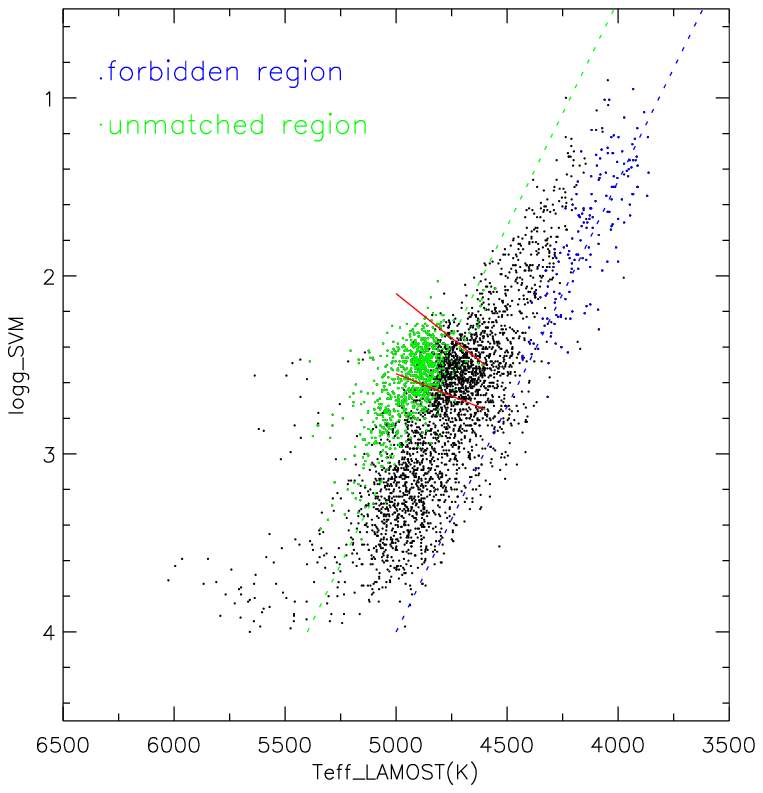}
\caption{The comparison of the $\teff$ versus $\logg$ diagram based on
new gravities from \cite{Liu15}. The symbols are the same as in the left-upper
panel of Fig.~4.}
\end{figure}

\begin{figure}[ht]
\includegraphics[width=14.5cm, angle=0]{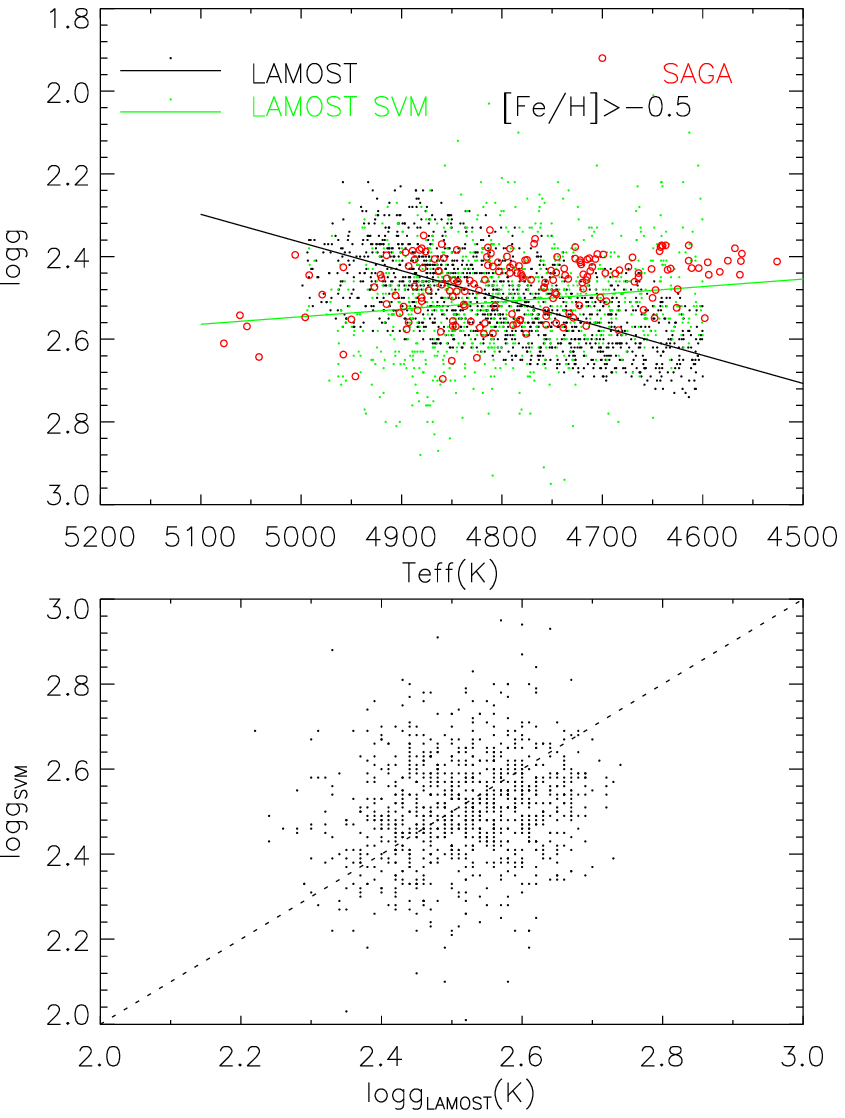}
\caption{The slope comparison in the $\teff$ versus $\logg$ diagram for RC stars
among new gravities from \cite{Liu15}, the LAMOST dataset and the SAGA survey by \cite{Casagrande14}.}
\end{figure}

\section{Summary}
We have compared the stellar parameters, $\teff$, $\logg$ and $\feh$,
between the LAMOST and the APOGEE datasets. We have identified the main
sequence of RC stars in the  $\teff$ versus $\logg$ diagram of the
LAMOST dataset, which behaves in a more reasonable way than that
in the APOGEE datasets. For RGB stars, the LAMOST dataset spans
a wider range than the APOGEE dataset in the  $\teff$ versus $\logg$ diagram,
and a group of stars with  $2.6>\logg>0.00253*\teff-8.67$ locates
at a forbidden region where no theoretical model predicts such kinds
of stellar parameters. We further exclude stars with their metallicity
of $\feh>-1$ unmatched with their positions in the  $\teff$ versus $\logg$ diagram
(outside the green dash line where RGB stars with $\feh<-1$ locate).

We have established a good metallicity calibration of $\feha=1.18*\fehl+0.11$
 ($\sigma=0.08$) for both RC and RGB stars, and a temperature calibration
of $\teffa=0.95*\teffl+210 (\sigma=53\,K)$ in consistent with the one-to-one
line within the measured errors. There is no clear trend in gravity between
the LAMOST and the APOGEE datasets for RC stars, and we may prefer the
the LAMOST dataset rather than the APOGEE dataset since the former
follows the general trend of increasing $\logg$ with decreasing $\teff$,
which is the RC feature as seen in the CMD of local RC stars in the field
and some old open clusters. For example, the CMD of an old open
cluster NGC6819 in \cite{Lee-Brown15} shows that $V$ magnitude
becomes fainter (corresponding to the decrease of luminosity)
 as $(B-V)$ color goes redder (indicating the decreasing temperature).
However, the RC sample of Casagrande et al. (2014) 
do not show this dependence, and
we need further study to clarify if the dependence of $\logg$
with $\teff$ for RC stars is true.
For RGB stars, we prefer calibrations of 
$\logga=0.000874*\teffl+0.588*\loggl-3.117$ ($\sigma=0.167$)
for  $\feh<-1$ and $\logga=0.000615*\teffl+0.697*\loggl-2.208 (\sigma=0.150)$ for $\feh>-1$ 
after
excluding stars in the forbidden region and the unmatched region
of the  $\teff$ versus $\logg$ diagram. Finally, we have found that
new gravities from the SVM method based on the seismic $\logg$ by \cite{Liu15} 
are more reliable than the original values in the LAMOST database
for stars in both the forbidden and the unmatched regions.

\begin{acknowledgements}
This study is supported by the National Natural Science
Foundation of China under grant Nos. 11222326, 11233004, 11390371, the
Strategic Priority Research Program of the Chinese Academy of Sciences Grant No. XDB01020300
and the National Key Basic Research Program of China (973 program)
Nos. 2014CB845701/703.\\
Guoshoujing Telescope (the Large Sky Area Multi-Object Fiber Spectroscopic Telescope LAMOST) is a National Major Scientific Project built by the Chinese Academy of Sciences. Funding for the project has been provided by the National Development and Reform Commission. LAMOST is operated and managed by the National Astronomical Observatories, Chinese Academy of Sciences.\\
Funding for SDSS-III has been provided by the Alfred P. Sloan Foundation,
the Participating Institutions, the National Science Foundation, and the U.S. Department
of Energy Office of Science. The SDSS-III web site is http://www.sdss3.org/.
SDSS-III is managed by the Astrophysical Research Consortium for the Participating
Institutions of the SDSS-III Collaboration including the University of Arizona,
the Brazilian Participation Group, Brookhaven National Laboratory, University of
Cambridge, University of Florida, the French Participation Group, the German
Participation Group, the Instituto de Astrofisica de Canarias, the Michigan
State/Notre Dame/JINA Participation Group, Johns Hopkins University, Lawrence
Berkeley National Laboratory, Max Planck Institute for Astrophysics, New Mexico
State University, New York University, Ohio State University, Pennsylvania State
University, University of Portsmouth, Princeton University, the Spanish Participation
Group, University of Tokyo, University of Utah, Vanderbilt University, University
of Virginia, University of Washington, and Yale University.
\end{acknowledgements}

\end{document}